\documentclass[pre,groupedaddress,showkeys,showpacs,twocolumn]{revtex4}
\usepackage{amsmath}
\usepackage{graphics}
\usepackage{graphicx}
\usepackage{amsfonts}
\usepackage{amssymb}
\usepackage{dcolumn}
\usepackage{bm}
\begin{document}
\title{Normal contact and friction of rubber with model randomly rough surfaces}
\author{S. Yashima\textit{$^{a,d,\ddag}$}}
\author{V. Romero\textit{$^{b,c,\ddag}$}}
\author{E. Wandersman\textit{$^{b,c}$}}
\author{C. Fr{\'e}tigny\textit{$^{a}$}}
\author{M.K. Chaudhury\textit{$^{e}$}}
\author{A. Chateauminois\textit{$^{a}$}}
\author{A. M. Prevost\textit{$^{b,c,\ast}$}}
\email[]{alexis.prevost@upmc.fr}
\affiliation{\textit{$^{a}$~Soft Matter Science and Engineering Laboratory (SIMM), CNRS / UPMC Univ Paris 6, UMR 7615, ESPCI, F-75005 Paris, France}}
\affiliation{\textit{$^{b}$~CNRS, UMR 8237, Laboratoire Jean Perrin (LJP), F-75005, Paris, France. E-mail: alexis.prevost@upmc.fr}}
\affiliation{\textit{$^{c}$~Sorbonne Universit\'es, UPMC Univ Paris 06, UMR 8237, Laboratoire Jean Perrin, F-75005, Paris, France}}
\affiliation{\textit{$^{d}$~Laboratory of Soft and Wet Matter, Graduate School of Life Science, Hokkaido Univ, Sapporo, Japan}}
\affiliation{\textit{$^{e}$~Department of Chemical Engineering, Lehigh University, Bethlehem  PA  18015, USA}}
\begin{abstract}
We report on normal contact and friction measurements of model multicontact interfaces formed between smooth surfaces and substrates textured with a statistical distribution of spherical micro-asperities. Contacts are either formed between a rigid textured lens and a smooth rubber, or a flat textured rubber and a smooth rigid lens. Measurements of the real area of contact $A$ versus normal load $P$ are performed by imaging the light transmitted at the microcontacts. For both interfaces, $A(P)$ is found to be sub-linear with a power law behavior. Comparison to two multi-asperity contact models, which extend Greenwood-Williamson (J. Greenwood, J. Williamson, \textit{Proc. Royal Soc. London Ser. A} \textbf{295}, 300 (1966)) model by taking into account the elastic interaction between asperities at different length scales, is performed, and allows their validation for the first time. We find that long range elastic interactions arising from the curvature of the nominal surfaces are the main 
source of the non-linearity of $A(P)$. At a shorter range, and except for very low pressures, the pressure dependence of both density and area of micro-contacts remains well described by Greenwood-Williamson's model, which neglects any interaction between asperities. In addition, in steady sliding, friction measurements reveal that the mean shear stress at the scale of the asperities is systematically larger than that found for a macroscopic contact between a smooth lens and a rubber. This suggests that frictional stresses measured at macroscopic length scales may not be simply transposed to microscopic multicontact interfaces.
\end{abstract}
\maketitle
%
\section*{Introduction}
\indent Surface roughness has long been recognized as a key issue in understanding solid friction between macroscopic bodies. As pointed out by the pioneering work of Bowden and Tabor~\cite{bowden1958}, friction between rough surfaces involves shearing of myriads of micro-asperity contacts of characteristic length scales distributed over orders of magnitude. The statistical averaging of the contributions of individual micro-asperity contacts to friction remains an open issue which largely relies on the contact mechanics description of multicontact interfaces. In early multi-asperities contact models such as the seminal Greenwood-Williamson's model (GW)~\cite{Greenwood1966}, randomly rough surfaces are often assimilated to a height distribution of non interacting spherical asperities which obey locally Hertzian contact behavior. Along these guidelines, some early models also attempted to describe the fractal nature of surface roughness by considering hierarchical distributions of asperities~\cite{archard1957}.
More refined exact elastic contact mechanics theories were also developped by Westergard \cite{Westergaard1939}, Johnson~\cite{johnson1985b} and Manners~\cite{manners2003,manners1998}, amongst others, in order to solve the problem of elastic contacts between one dimensional periodic wavy surfaces. Most of the subsequent generalisations of elastic contact theories to randomly rough surfaces are more or less based on a spectral description of surface topography~\cite{persson2001,hyun2004,campana2008,Hyun2007}. Within the framework of linear (visco)elasticity or elasto-plastic behavior, these theories allow estimation of the pressure dependence of the distribution of microcontacts size and pressure at various length scales. From an experimental perspective, elucidation and validation of these models using microscopic randomly rough surfaces such as abraded or bead blasted surfaces is compromised by the difficulties in the measurement of the actual distribution of microcontact areas at the micrometer scale. 
Although early attempts were made by Dieterich and Kilgore~\cite{dieterich1996} with roughened surfaces of transparent materials using contact imaging techniques, direct comparison of the experimental data with contact mechanics models lacks clarity.\\
\indent In this study, we take advantage of recent advances in sol-gel and micro-milling techniques to engineer two types of model randomly rough and transparent surfaces with topographical characteristics compatible with GW's model of rough surfaces \cite{Greenwood1966}. They both consist of statistical distributions of spherical asperities whose sizes ($\sim 20$ $\mu$m up to  200 $\mu$m) allow for an optical measurement of the spatial distributions of the microcontacts areas. In their spirit, these experiments are along the line of Archard's previous investigations~\cite{archard1957}, which used model perspex surfaces consisting of millimeter sized spherical asperities of equal height. However, in Archard's investigations, a small number of asperities were used. Furthermore, technical limitations in the estimation of variation of heights of asperities did not allow for a statistical analysis of the load dependence of the distributions of microcontact areas. Here, using a sphere-on-plane contact geometry 
with different statistical distributions of micro-asperities, we probe the elastic interactions between asperities (see \textit{e.g.}~\cite{greenwood1967,ciavarella2006,ciavarella2008,guidoni2010}) by directly comparing the measured distributions of the real area of contact to the predictions of two different multi-asperity contact models. We show how the use of textured surfaces allows an accurate validation of these models that permits an investigation of the statistical distribution of contact pressure, number of microcontacts and microcontact radii distributions. In the last part of the paper, we present the results of a preliminary study that illustrates how such model systems can be used to investigate the relationship between frictional properties and real contact areas.\\
\section*{Materials and Techniques}
Two types of randomly rough surfaces covered with spherical caps were designed using two different techniques as described below. The first surface (RA for Rigid Asperities) consists of glass lenses (BK7, Melles-Griot, radius of curvature 13 mm) covered with a distribution of micrometer sized rigid asperities with varying heights and radii of curvature. The second surface (SA for Soft Asperities) is made of a nominally flat silicone slab decorated with a random spatial distribution of soft spherical micro-asperities with equal radius of curvature and varying heights.
\begin{figure}[ht]
	\centering
	\includegraphics[width=\columnwidth]{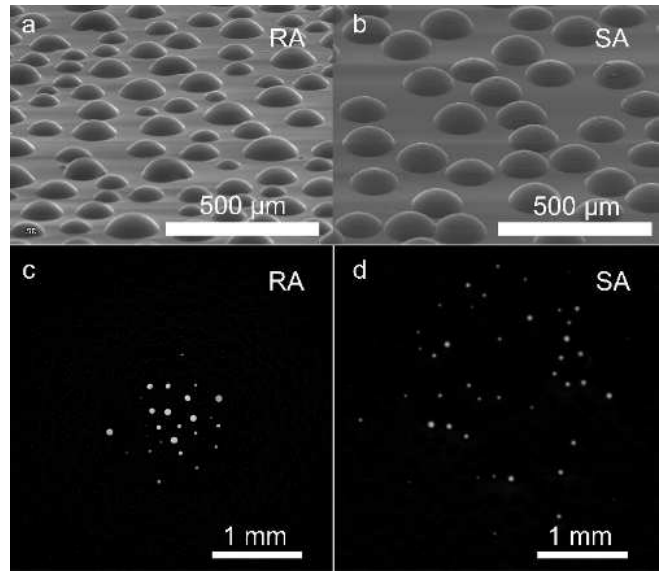}
	\caption{(a) SEM image topography of a RA$^+$ sol-gel replica ($\phi=0.41$). (b) Same with an SA PDMS replica of a micro-milled mold ($\phi=0.4$). (c) microcontacts spatial distribution with RA$^+$ ($P = 22$~mN). (d) Same with the SA of (b) and a lens of radius of curvature 128.8 mm ($P = 20$~mN). (c-d) are image differences with a reference non-contact image. Note the size difference in the apparent contact related to the difference in curvature of both indenters.}
	\label{fig:sem} 
\end{figure}
\subsection*{RA lenses}
\begin{figure}
	\includegraphics[width=\columnwidth]{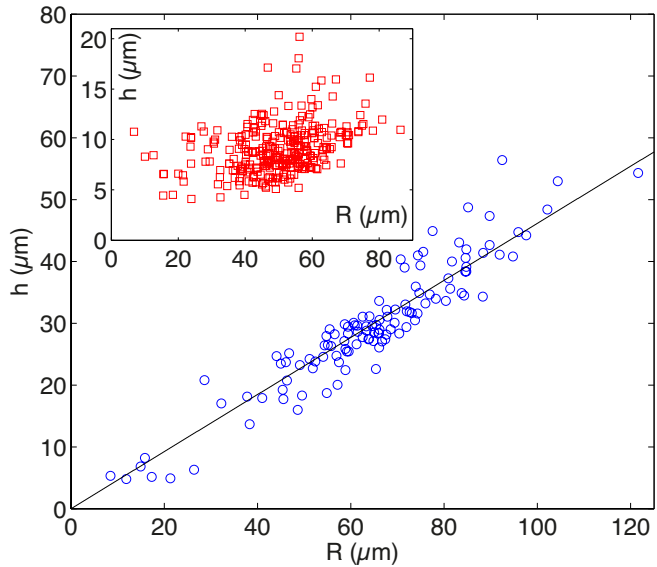}
	\caption{(Color online) Height $h$ of the spherical micro-asperities as a function of their radius of curvature $R$ for the RA$^+$ lens ($\phi=0.41$). Inset: Same for the RA$^-$ lens ($\phi=0.34$). The solid line is a linear fit of the data.}
	\label{fig:h_vs_R} 
\end{figure}

RA's topography was obtained by replicating condensed liquid droplets on a hydrophobic surface. Water evaporating from a bath heated at 70$^\circ$C  was first allowed to condense on a HexaMethylDiSilazane (HMDS) treated hydrophobic glass slide kept at room temperature, resulting in a surface with myriads of droplets. This surface was then covered with a degassed mixture of a PolyDiMethylSiloxane cross-linkable liquid silicone (PDMS, Sylgard 184, Dow Corning) cured at 70$^\circ$C for 2 hours. One is left, upon demolding, with a PDMS surface with concave depressions, which are negative images of the condensed water droplets. These PDMS samples then serve as molds to replicate rigid equivalents on the glass lenses using a sol-gel imprinting process fully described elsewhere~\cite{letailleur2010}. An example of the resulting pattern with smooth spherical caps of various sizes is shown in Fig.~\ref{fig:sem}a. By changing the time of exposure $t_{exp}$ of the HMDS treated glass to water vapor, different surfaces 
with different asperity sizes and densities are obtained as a result of droplet coalescence during the water condensation process. Two patterns with small (\textit{resp.} large) asperities were made with $t_{exp}=15$~s (\textit{resp.}  60~s). They are respectively referred to as RA$^-$ and RA$^+$. Their topography at the apex was characterized with an optical profilometer (Microsurf 3D, Fogale Nanotech) to extract the mean surface fraction $\phi$ covered by the asperities (Table 1) and the distributions of their heights $h$ and radii of curvature $R$. Both distributions are found to be Gaussian (not shown) with means $\bar h$, $\bar R$ and standard deviations given in Table 1. For RA$^+$, $h$ is found to be proportional to $R$ (Fig.~\ref{fig:h_vs_R}). This suggests that the spherical shape of the asperities is uniquely controlled by the contact angle $\theta$ of water droplets on the HMDS treated surface prior to molding. In this case, one expects, indeed, the relationship $h=R(1-\cos \theta)$. Fitting the 
data of Fig.~\ref{fig:h_vs_R} yields $\theta~\sim$~57$^\circ$, very close to 55$^\circ$ which is the value of the advancing contact angle we measured for water droplets on HMDS treated glass. For RA$^-$ however, no evident correlation has been observed, for which we have no clear explanation (Fig.~\ref{fig:h_vs_R}, inset).\\
\begin{table}[ht]
	\caption{\label{tab:table1}
		RA's mean topographical characteristics}
	\begin{tabular*}{0.48\textwidth}{@{\extracolsep{\fill}}lllll}
		\hline
		$t_{exp}$(s)&$\phi$&$\bar h$($\mu$m)&$\bar R$($\mu$m)\\
		\hline
		\hline
		15 & 0.34~$\pm$~0.02 & 9.0~$\pm$~2.4 & 49.6~$\pm$~12.8 $^{a}$ \\
		60 & 0.41~$\pm$~0.05 & 29.6~$\pm$~10.1 & 64.4~$\pm$~19.6 $^{b}$ \\
		\hline
	\end{tabular*}
	$^{a}$ from 293 asperities. \\
	$^{b}$ from 119 asperities.
\end{table}
\subsection*{SA samples}
SA samples were obtained by cross-linking PDMS in molds (2.5 mm deep) fabricated with a desktop CNC Mini-Mill machine (Minitech Machinary Corp., USA) using ball end mills of radius 100 $\mu$m, allowing to design, with 1 $\mu$m resolution, patterns with controlled surface densities and height distributions (Fig.~\ref{fig:sem}b). Spherical cavities were randomly distributed over 1 cm$^2$ with a non overlapping constraint with two different surface densities $\phi=0.1$ and $0.4$. Their  heights as obtained  from a uniform random distribution were in the range [30--60]~$\mu$m. SA samples with $\phi=0.1$ are thus referred to as SA$^-$ further down, and those with $\phi=0.4$ as SA$^+$. Half of the bottom of the mold was kept smooth so that SA samples had both a patterned part and a smooth one. The smooth part was used in a JKR contact configuration \cite{Johnson1985a}, which allowed measurement of each sample Young's modulus $E$. Secondly, it provided means to locate accurately the center of the apparent contacts 
formed on the 
patterned part. Since contacts with the patterned part were obtained by a simple translation of the sample, the center within the contact images was taken as the center of the JKR circular contact, obtained using standard image analysis.\\

\indent As detailed above, RA samples display spatial and height distributions of asperities set by both the evaporation and the sol-gel processes, which can only be characterized \textit{a posteriori}. SA samples however, have a statistical roughness which can be finely tuned with any desired pattern, both in height and spacing. As a result, SA flat surfaces are very appropriate for the statistical investigation of 
contact pressure distribution as they can be produced at centimeter scales thus allowing for several realizations of the contact at different positions on the patterned surface. Nevertheless, contrary to SA asperities which always present a microscopic surface roughness inherent to the milling procedure, RA micro-asperities are very smooth. It thus makes them especially suitable for the investigation of frictional properties, as microcontacts obtained with a smooth rubber substrate can be assimilated to single asperity contacts.
\subsection*{Experimental setups}
\indent For RA lenses, normal contact experiments were performed by pressing the lenses against a thick flat PDMS slab under a constant normal load $P$. Its thickness ($\sim15$~mm) was chosen to ensure semi-infinite contact conditions (\textit{i.e.} the ratio of the contact radius to the specimen thickness was more than ten~\cite{Gacoin2006}). For SA flat samples, sphere-on-plane contacts were obtained by pressing them against a clean BK7 spherical lens (LA1301, Thorlabs Inc.) with a radius of curvature of 128.8 mm, $\sim 10$ times larger than the radius of curvature of the patterned RA lenses. To ensure comparable semi-infinite contact conditions, SA samples remained in adhesive contact against a $\sim15$~mm thick PDMS slab. The experiments were performed with  a home made setup described in \cite{prevost2013,Romero2013}. Using a combination of cantilevers and capacitive displacement sensors, both the normal ($P$) and interfacial lateral ($Q$) forces are monitored in the range [0--2.5]~N with a resolution 
of $10^{-3}$~N. This setup also provides simultaneous imaging of the microcontacts with the combination of a high resolution CCD camera (Redlake ES2020M, 1600$\times$1200 pixels$^2$, 8 bits) and a long--working distance Navitar objective. Once illuminated in transmission with a white LED diffusive panel, microcontacts appear as bright disks. Measuring their areas using standard image thresholding techniques provides a direct measure of their entire spatial distribution. The total true area of contact $A$ is then obtained by summing all microcontact areas. In addition, assuming the validity of Hertzian contact theory at the scale of the asperity and knowing $E$, radii of curvature $R$ of each asperity and $\nu=0.5$ the Poisson's ratio \cite{prevost2013,Romero2013}, the disks radii $a_i$ are a direct measure of the local normal forces $p_i$ since
\begin{equation}
	p_i=\frac{4 E a_i^3}{3 (1-\nu^2) R} 
	\label{EqHertz}	
\end{equation}
As described previously \cite{Romero2013}, a linear relationship between the total normal load $P_c=\sum_{i} p_i$ and the measured $P$ is systematically found for all SA samples, thus validating Hertz assumption. However, the slope of $P_c$ versus $P$ depends slightly on the optical threshold used to detect $a_i$. To recover a unit slope, we thus calibrated the optical threshold with a reference sample of known Young's modulus. For all other samples, we then kept the same optical threshold and tuned $E$ for each sample within its measured uncertainties to recover a unit slope. Note that Hertz contact theory assumes that $a_i/R \ll 1$ in order to stay in the linear elastic range. In our experiments, we find that, at the highest normal load, $a_i/R$ is at maximum of the order of $0.3$. Investigations by Liu and coworkers~\cite{Briscoe1998} using micro-elastomeric spheres in contact with a plane (contact radius $a$) have shown however that Hertz theory remains accurate for values of $a/R$ up to $\sim 0.33$.\\
For RA samples, such a calibration method could not be applied as it requires knowing the radii of curvature of all asperities to evaluate $p_i$. Because of this limitation\footnote{Measurements of radii of curvature were performed using profilometry images obtained at a high magnification. Identifying for a given asperity its radius of curvature would imply matching the position of this asperity with its position in a zoomed out image of the macroscopic apparent contact.}, we chose the threshold arbitrarily from the contact images between their two extremal values for which the change in total area was found to vary marginally. Consequently, it was not possible to measure any local normal force distribution for RA samples.\\
\indent Friction experiments with RA patterned lenses were performed with another experimental setup described earlier~\cite{piccardo2013}. RA lenses were rubbed against a smooth PDMS slab ($E~=~3\pm0.1$~MPa) keeping both $P$ and the driving velocity $v$ constant. The setup allowed variation of $v$ from a few $\mu$m\,s$^{-1}$ up to 5~mm\,s$^{-1}$ thus allowing simultaneous measurements of $P$ and $Q$ with a resolution of $10^{-2}$~N. 
\section*{Multi-asperity contact models}
\begin{figure}
	\includegraphics[width=\columnwidth]{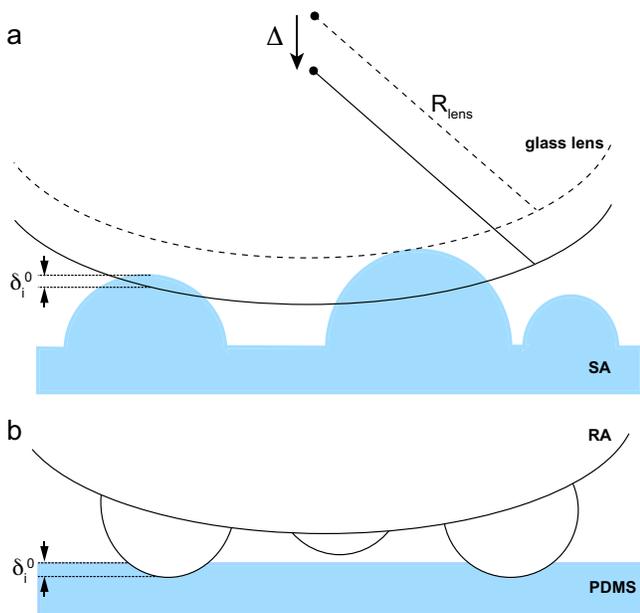}
	\caption{Sketch of the geometric configuration for the indentation of (a) SA and (b) RA surface topography. For both configurations, $\Delta$ is the prescribed indentation depth taking as a reference for the vertical position of the indenting sphere the altitude at which the smooth surface is touching the uppermost asperity.}
	\label{fig:contact_sketch} 
\end{figure}

To investigate quantitatively the effects of elastic interactions between micro-asperity contacts on the real contact area and related pressure distribution, two different multi-asperity contact models were considered, both of which include elastic interactions at different length scales. The first one was derived by Greenwood and Tripp (GT)~\cite{greenwood1967} as an extension to the case of rough spheres of the seminal model of Greenwood and Williamson (GW) for the contact between nominally flat surfaces. The second one was developed more recently by Ciavarella~\textit{et al.}~\cite{ciavarella2006,ciavarella2008}. It consists in a modified form of GW's model, with elastic interactions between microcontacts incorporated in a first order-sense. Both models describe the contact mechanics of rough surfaces with random distributions of spherical asperities, which is what we investigate here experimentally. As a consequence of this simplified form of surface topography, it was not necessary to consider more 
refined contact models based on a spectral description of the surfaces such as Persson's model~\cite{persson2001}.\\
\indent In GT's model, Hertz theory of elastic contact between a smooth sphere and a smooth plane is extended by adding roughness to the plane. As a starting point, the relationship between the local pressure and the local real contact area within an elementary portion of the rough contact is assumed to obey GW's theory. Accordingly, micro-asperity contacts are supposed to be Hertzian and to be independent, that is, the elastic displacements due to the normal force exerted on one asperity has negligible effect on any other asperity. However, use of GW's relationship requires that the separation of both surfaces at any location within the macroscopic contact is known, \textit{i.e.} that the shape of nominal surfaces under deformation is determined. This requirement is deduced from linear elasticity theory (Green's tensor, see reference~\cite{landau1986} for instance) that introduces long range elastic interactions at the scale of the apparent Hertzian contact. As opposed to GW's model, which can be derived 
analytically, in GT's model, calculation of the real contact area and pressure distribution can only be done with an iterative numerical integration of a set of coupled equations, as described in~\cite{greenwood1967}.\\
\indent In Ciaravella \textit{et al.}'s model, the approach includes in the first order-sense elastic interactions between Hertzian micro-asperity contacts, \textit{i.e.} for every asperity a displacement is imposed which is sensitive to the effect of the spatial distribution of Hertzian pressures in the neighboring asperities. For each micro-asperity contact, a shift of the position of the deformable surface is introduced, which results from the vertical displacement caused by the neighboring ones. Accordingly, the indentation depth $\delta_i$ of the $i^\text{th}$ micro-asperity contact is
\begin{equation}
	\delta_i = \delta_i^0 + \sum_{j\neq i}^N\alpha_{ij}\delta_j^{3/2}\:,
	\label{eq:delta}
\end{equation}
where $\delta_i^0>0$ is the indentation depth in the absence of any elastic coupling between microcontacts, and $\alpha_{ij}$ are the elements of the interaction matrix. As shown in Fig.~\ref{fig:contact_sketch}, $\delta_i^0$ is a purely geometrical term simply given by the difference between the positions of the two undeformed surfaces for the prescribed indentation depth $\Delta$. The sum in the rhs of eqn~(\ref{eq:delta}) represents the interaction term derived from Hertz contact theory. Our study slightly differs from Ciavarella~\textit{et al.}'s model as we take for $\alpha_{ij}$ an asymptotic expansion of the Hertz solution for the vertical displacement of the surface, instead of its exact expression. Elements $\alpha_{ij}$ of the interaction matrix thus read
\begin{equation}
	[\alpha_{ij}] = - \frac{4 \sqrt{R_j}}{3 \pi} \frac{1}{r_{ij}} \:\:, i \neq  j\:,
	\label{eq:matA}
\end{equation}
where $r_{ij}$ is the distance between asperities $i$ and $j$ and $R_j$ is the radius of curvature of the $j^\text{th}$ asperity. This approximation avoids evaluating at each step of the calculation the interaction matrix $[\alpha_{ij}]$, which consequently depends only on the surface topography. Such an approximation is valid as long as the average distance between asperities $L$ is much larger than the average asperity microcontact radius $a$. For RA samples, optical measurements reveal that this criterion is satisfied as the ratio $L/a$, which is a decreasing function of $P$, remains between 6 and 8. For SA samples, one also measures that $L/a \approx 16-32$ for SA$^-$ and $L/a \approx 9-15$ for SA$^+$. The above detailed models are obviously valid as long as no contact occurs in regions between the top parts of the spherical caps.\\
%
%
\section*{Normal contacts}
\subsection*{RA measurements}
\begin{figure}
	\includegraphics[width=\columnwidth]{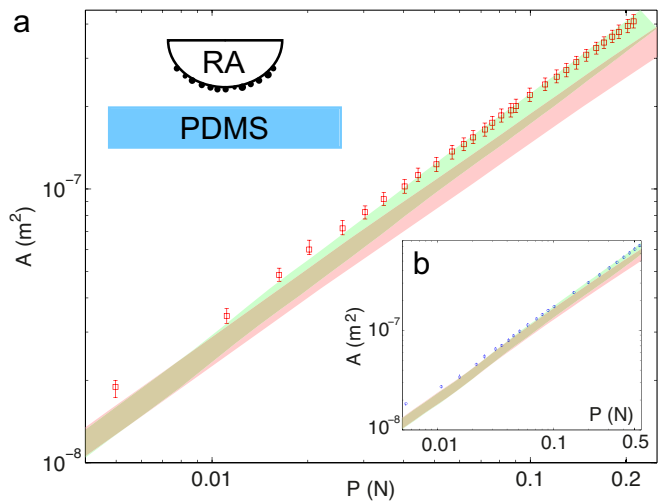}
	\caption{Log-log plot of the real area of contact $A$ versus $P$ for both RA$^{-}$ (a) and RA$^{+}$ (b) lenses. The upper and lower limits of the error bars correspond to the total areas measured with the arbitrarily chosen extremal values of the optical threshold (see text). Red shaded areas correspond to the predictions of Ciavarella~\textit{et al.}'s model~\cite{ciavarella2006,ciavarella2008} by setting $\alpha_{ij}$ to 0 in eqn~(\ref{eq:delta}). Green areas correspond to $\alpha_{ij} \neq 0$. Areas extent characterizes the scatter in the simulations, arising from uncertainties in the experimental determination of the topography parameters.}
	\label{fig:contact_area} 
\end{figure}

In order to stay consistent with the hypothesis of the contact models, true contact area measurements for RA lenses were performed for normal loads $P$ for which only tops of the micro-asperities make contact with the PDMS slab. While for RA$^+$ lenses, this is observed for the entire range (up to 0.6~N) of $P$, for RA$^-$ lenses this occurs as long as $P \leq 0.2$~N. Figure~\ref{fig:contact_area} shows the total contact area $A$ versus $P$ for both RA lenses contacting a smooth PDMS substrate. $A(P)$ exhibits a non-linear power law behavior with the following exponents: $0.812 \pm 0.009$ for RA$^-$ and $0.737 \pm 0.042$ for RA$^+$.\\ 
To compare these results with Ciaravella \textit{et al.}'s model, calculations were carried out using simulated lens topographies generated from Gaussian sets of asperity heights calculated using the experimental parameters reported in Table~\ref{tab:table1}. The radii of curvature of the asperities were varied as a function of their heights using the experimentally measured $R(h)$ relationship. Asperities were spatially distributed according to a uniform distribution with a non-overlap constraint. In order to minimize bias in their spatial distribution, asperities were positioned by decreasing size order.\\
Figure~\ref{fig:contact_area} shows the results of such simulations using Ciavarella's model. Uncertainties in the experimental determination of surface parameters (mainly the $R(h)$ relationship) were found to result in some scatter in the simulated $A(P)$ response. In order to account for this scatter, the simulated curves are represented as colored areas in Fig.~\ref{fig:contact_area}. A good agreement is observed between theory and experiments only when elastic interactions are accounted for. Without such interactions (\textit{i.e.} when the term $\alpha_{ij}$ in eqn~(\ref{eq:delta}) is set to zero), the actual contact area at a given $P$ is clearly underestimated.\\
\subsection*{SA measurements}
\begin{figure}
	\includegraphics[width=\columnwidth]{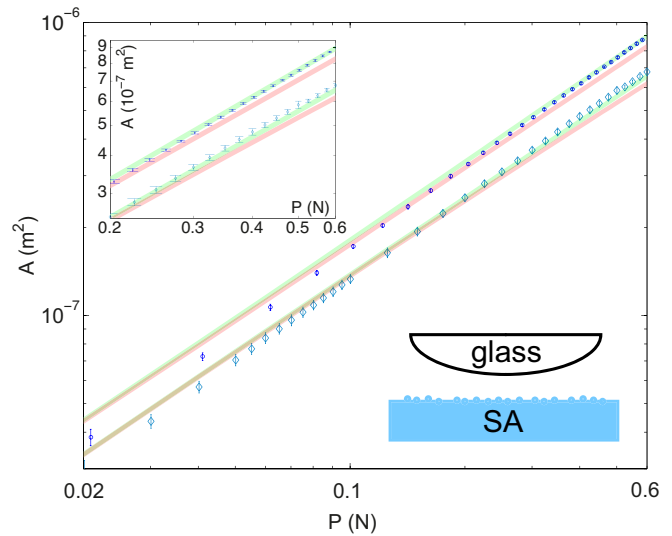}
	\caption{Log-log plot of the real area of contact $A$ versus $P$ for both SA$^{-}$ ($\phi=0.1$, blue diamonds) and SA$^{+}$ ($\phi=0.4$, blue circles) samples. The inset is a close up for $0.2~\leq~P~\leq~0.6$~N. Error bars are given by the standard deviation of $A$ on 24 different contact configurations. Red shaded areas correspond to the predictions of Ciavarella~\textit{et al.}'s model~\cite{ciavarella2006,ciavarella2008} by setting $\alpha_{ij}$ to 0 in eqn~(\ref{eq:delta}). Green areas correspond to $\alpha_{ij} \neq 0$. Areas extent characterizes the scatter in the simulations, arising from uncertainties in the experimental determination of $E$.}
	\label{fig:contact_areaSA} 
\end{figure}

For SA samples in contact with the glass lens of radius of curvature 128.8 mm, microcontacts always occur at the top of the asperities for the whole investigated $P$ range up to 0.6~N. For each $P$, the real area of contact $A$ was averaged over $N=24$ different locations on the sample. This allowed us to probe statistically different contact configurations while reducing the error on $A$ by a factor $\sqrt{N}$. Figure~\ref{fig:contact_areaSA} shows the resulting $A$ versus $P$ for both SA$^-$ and SA$^+$ samples. As found with RA lenses, $A(P)$ curves are also sub-linear and are well fitted by power laws. For both tested surface densities, power law exponents are found to be density independent, with $0.945 \pm 0.014$ for SA$^-$ and $0.941 \pm 0.005$ for SA$^+$. Changing $\phi$ from 0.1 to 0.4 mainly results in an increase of $A(P)$ at all $P$ (Fig.~\ref{fig:contact_areaSA}). As previously done with RA samples, both SA data sets are compared 
to Ciaravella \textit{et al.}'s model~\cite{ciavarella2006,ciavarella2008} predictions, with both $\alpha_{ij}=0$ and $\alpha_{ij} \neq 0$. Calculations were performed using the exact topography used to make SA samples, and $A$ versus $P$ curves were obtained with the exact same 24 contact configurations. Errors on the calculated $A$ values were obtained by varying Young's modulus within its experimental uncertainties, yielding the shaded areas of Fig.~\ref{fig:contact_areaSA}. Red shaded areas correspond to $\alpha_{ij}$ to 0 in eqn~(\ref{eq:delta}), while green areas correspond to $\alpha_{ij} \neq 0$. At low normal loads ($P~\le~0.1$~N), the effect of the elastic interaction on $A$ is almost negligible, but it becomes more pronounced at higher ones ($P~>0.1$~N), resulting in a larger true contact $A$. As shown on Fig.~\ref{fig:contact_areaSA}, our data at $P~>0.1$~N is clearly better captured by the interacting model rather than the non-interacting one for both surface densities.\\

These $A(P)$ measurements, together with those obtained with RA lenses, indicate that including an elastic interaction is thus essential to have a complete description of the contact mechanics of such systems. Yet, it remains unclear which of the short range (interaction between neighboring asperities) and/or long range (determined by the geometry of the macroscopic contact) parts of the elastic interaction predominate. We now address precisely this question in the following.\\
\subsection*{Role of elastic interactions}
\begin{figure*}[ht]
	\centering
	\includegraphics[width=\textwidth]{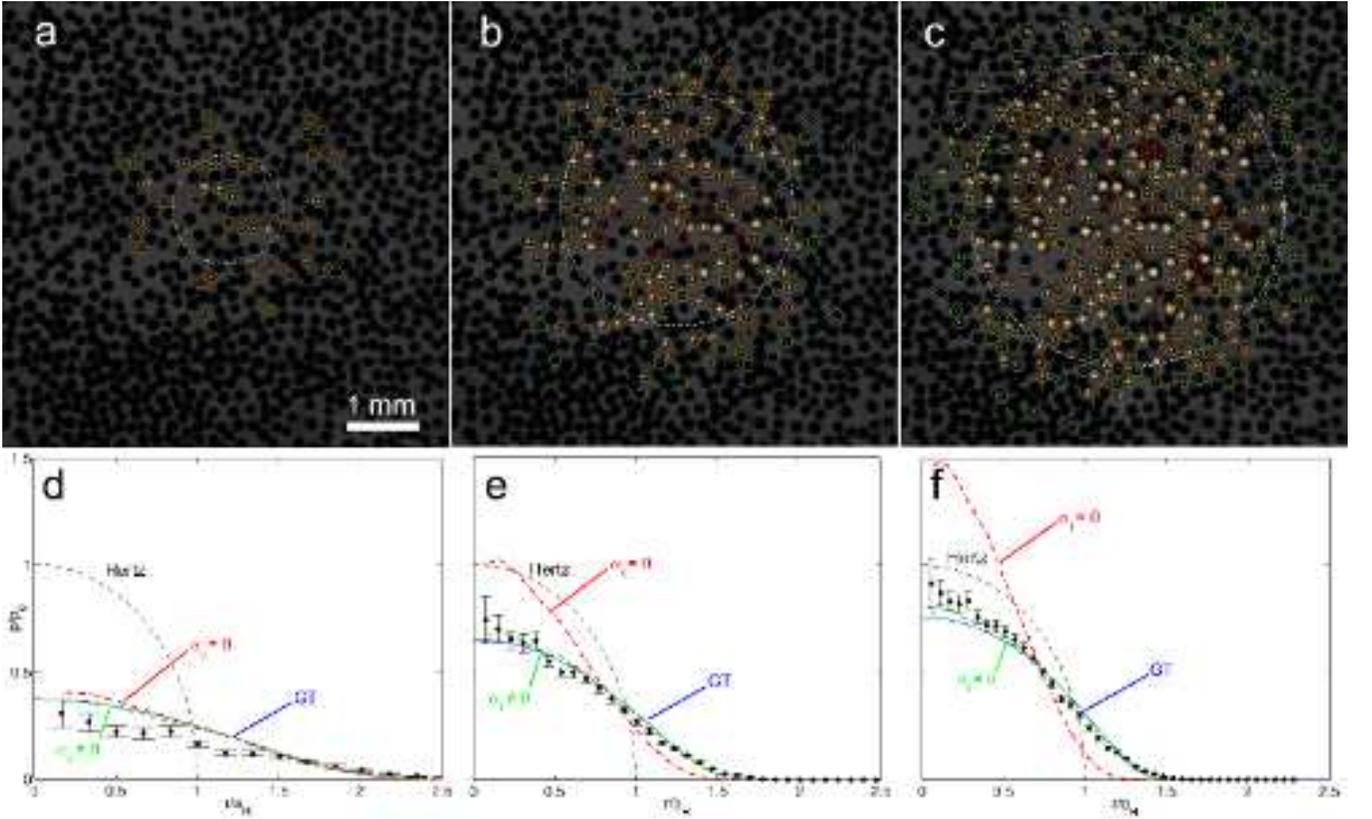}
	\caption{(Color online) (a), (b), (c) Images of the interface at $P=0.02, 0.2, 0.5~$N with the $\phi=0.4$ SA sample. microcontacts appear as the white disks. Green (\textit{resp.} red) circles indicate Ciaravella \textit{et al.}'s model predicted microcontacts with $\alpha_{ij} \neq 0$ (\textit{resp.} $\alpha_{ij} = 0$). On all images, the white dashed line circles delimit Hertz contacts for the corresponding $P$. (d),(e),(f) Angularly averaged pressure $p$ distribution as a function of the distance to the center $r$ on a SA sample with $\phi = 0.4$ at increasing normal loads $P$. Both $p$ and $r$ are normalized by respectively, Hertz' maximum pressure $p_0$ and Hertz contact radius $a_H$. The black dashed line corresponds to Hertz prediction. Blue solid lines are fits using Greenwood-Tripp model (GT) with a uniform asperity height density and same surface density $\phi$. The red dot-dashed lines are predictions of Ciaravella \textit{et al.}'s model~\cite{ciavarella2006,ciavarella2008} setting the 
		interaction term $\alpha_{ij}=0$, while the green dashed lines correspond to the full model with $\alpha_{ij} \neq 0$. Both latter predictions are statistical averages over 1000 independent pattern realizations with $\phi=0.4$ and a uniform height distribution.}
	\label{fig:contact_area_pprofiles} 
\end{figure*}
\textit{True contact area load dependence}\\
\indent Using contact imaging techniques, we were able to probe how the total true contact area varies with the applied load for contacts between a smooth surface and the different model rough surfaces decorated with spherical caps. For all sizes and spatial distributions of the micro-asperities tested here, we found that $A(P)$ curves could be satisfactorily described within the framework of a simple rough contact model with a classical assumption that Hertzian contact occurs at the scale of the micro-asperities. As opposed to both GW's and GT's models, our approach takes into account in an approximate manner the elastic coupling between asperities which is often neglected to fully describe the contact mechanics of rough interfaces.\\
For all investigated SA topographies, a nearly linear relationship is found for $A(P)$, which is consistent with the conclusions of the paper of Greenwood and Tripp\cite{greenwood1967} that states that $A(P)$ is ''approximately'' linear. More generally, our findings for SA surfaces do not depart from most of asymptotic development at low $P$ of most current rough contact models for nominally flat surfaces \cite{campana2008}. Such models, indeed, also predict a linear $A(P)$ relationship. Conversely, for RA topographies, a non-linear power law like $A(P)$ relationship is found. Such deviations from linearity was actually pointed out in recent theoretical works by Carbone and Bottiglione~\cite{carbone2008} for nominally plane--plane rough contacts. These authors pointed out indeed that asperity contact models deviate very rapidly from the asymptotic linear relation even for very small, and in many cases, unrealistic vanishing applied loads. For our present sphere--on--plane contact, it is legitimate to wonder 
if the 
magnitude of the deviations arises either from the differences in the asperities height and size distributions and/or the macroscopic curvatures of the spherical indenter. To provide an answer to this question, simulations using Ciaravella's \textit{et al.}'s model, with the exact same asperities distribution (height, radius of curvature and lateral distribution) but different radii of curvature $R_l$ of the macroscopic lens indenter ($R_l=13$~mm and $R_l=128.8$~mm, as in the experiments) were performed. In both cases, $A(P)$ curves are found to follow asymptotically (for $0.005 \le P \le 1$N) a power law, whose exponent is $\sim 0.86$ with $R_l=13~$mm and $\sim 0.93$ with $R_l=128.8$~mm. Decreasing $R_l$ thus enhances the nonlinearity of the $A(P)$ relationship. It is likely that such effects simply result from the fact that the increase in the gap between both the PDMS and the lens from the edges of the contact is larger for a lens with a small radius of curvature. For a load increase $\delta P$, the 
increase in the 
number of microcontacts at the periphery of the apparent contact area is thus expected to be more pronounced with a large $R_l$. This should translate into a more linear $A(P)$ dependence for large $R_l$. This hypothesis is further supported by a simple calculation detailed in Appendix A. Assuming that the rough contact obeys Hertz law at the macroscopic length scale, one can express the gap height between surfaces at the periphery of the contact as a function of the Hertzian radius and the radius of curvature of the indenting lens. Equating this gap height to the standard deviation of the height distribution yields a characteristic length scale $\Delta$ which corresponds to the size of the annular region surrounding the Hertzian contact. This length is found to vary as $\Delta \propto R_l^{5/9}P^{-1/9}$. This confirms that for a given applied load, the extension of the contact area from its Hertzian value, as resulting from microasperities contacts, should be enhanced when $R_l$ increases.\\
Of course, it is expected that the non-linearity of the $A(P)$ relationship could also depend on the statistical properties of the asperity distributions. This is indeed suggested by eqn.~(\ref{eq:deltac}) which predicts that $\Delta$ scales as $\sigma^{2/3}$, where $\sigma$ is the standard deviation of the height distribution of asperities. One can also mention the early theoretical work of Archard~\cite{archard1957}, based on hierarchical distribution of spherical asperities on a spherical indenter. This model predicts that $A(P)$ follows a power law whose exponent varies between 2/3 (\textit{i.e.} the limit of the smooth Hertzian contact) and unity (when the number of hierarchical levels of asperities is increased).\\
Before addressing further the issue of the elastic interactions between microcontacts, some preliminary comments are warranted, regarding the sensitivity of the $A(P)$ relationship to the details of the spatial distribution of microasperities. For that purpose, one can consider a comparison between experimental and theoretical results for RA patterns. While the micro-asperities were distributed spatially according to a uniform random distribution in the simulations, such a distribution probably does not reproduce very accurately the features of the droplet pattern. As a result of droplet coalescence during condensation, some short distance order is probably achieved between asperities as suggested by a close examination of Fig.~\ref{fig:sem}a. However, the good agreement between the experiments and the simulations in Fig.~\ref{fig:contact_area}a shows that the load dependence of the actual contact area is not very sensitive to the details in the spatial distribution of asperities. As far as the normal load 
dependence of the real contact area is considered, the relevant features of surface topography are thus likely to be mainly the surface density of micro-asperities, and their size and height distributions.\\
\textit{Microcontacts and pressure spatial distributions}\\
\indent So far, we only considered the effect of the elastic interaction on the load dependence of $A$, and thus neglected any spatial dependence of the microcontacts distribution. Direct comparison of such data with Ciaravella \textit{et al.}'s model calculations is not easily accessible for RA samples since it would require a knowledge of all asperities positions and respective radii of curvature. With SA samples however, this can be easily done, as positions and radii of curvature of asperities are known by design of the micromilled pattern. Figures~\ref{fig:contact_area_pprofiles}a-b-c show such direct comparison at three increasing normal loads $P$ ($P=0.02, 0.2, 0.5~$N) for the case of the SA$^+$ sample. As expected, predicted microcontacts with $\alpha_{ij} \neq 0$ almost always match the measured microcontacts (see the green circles on the figure). For comparison, red circles at the predicted positions of the model without elastic interaction have been overlapped on the contact images. Clearly, the 
non-interacting model predicts contacts at locations within the apparent contact which are not seen in the experiment.\\
\indent To perform a more quantitative comparison with theoretical predictions, we computed for both the experimental and calculated points, the local radial pressure profiles $p(r)$. The latter, which is expected to be radially symmetric for a sphere--on--plane normal contact, was obtained by summing up local forces $p_i$ exerted on all microcontacts located within an annulus of width $dr=0.25$~mm and radius $r$ centered on the apparent contact center (obtained from JKR experiments). To reduce the statistical error, averaging of $p(r)$ for several contact configurations was then performed. For the experiment, 24 contact configurations (compatible with the size of the SA pattern) at different locations on the same SA pattern were used. For the calculated data (Ciaravella \textit{et al.}'s model), 1000 statistically different SA patterns were used and normal loading was done at the center of the SA pattern. Both $\alpha_{ij}=0$ and $\alpha_{ij} \neq 0$ data were computed. To test the effect of including an 
elastic interaction at different length scales, we also computed $p(r)$ as predicted by GT's model. As discussed earlier, this model indeed constitutes in some sense a 'zeroth order approximation' of Ciaravella \textit{et al.}'s model, as it only takes into account long range elastic interactions whose extent is set by the size of the apparent contact. GT's calculation was implemented with Mathematica 9 (Wolfram Research Inc., USA), using a random asperities height distribution with heights chosen uniformly between 30 and 60 $\mu$m.\\
\indent Figures~\ref{fig:contact_area_pprofiles}d--e--f show the results on the example of SA$^+$ for the three increasing loads $P$ of Figs.~\ref{fig:contact_area_pprofiles}a--b--c. As already anticipated from Figs.~\ref{fig:contact_area_pprofiles}a--b--c, Ciaravella \textit{et al.}'s model with $\alpha_{ij} \neq 0$ gives a reasonably good fit of the measured data. Taking $\alpha_{ij} = 0$ yields larger discrepancy with the experimental points, revealing that, on average, the effect of the elastic interaction is to increase significantly the apparent radius of contact, the higher the normal load $P$. As pointed out by Greenwoood and Tripp in their original paper, the effect of roughness is to add a small tail to the Hertzian pressure distribution which corresponds to the annular region around the Hertzian contact in which the separation is comparable with the surface roughness. Indeed, as already mentioned earlier, an order of magnitude of this tail is provided by the characteristic length $\Delta$ which 
scales as $R^{5/9}\sigma^{2/3}$ (see Appendix A). It can be noted that this scaling is very close to that deduced from different arguments by Greenwood and Tripp (\textit{i.e.} $\Delta \propto \sqrt{R\sigma}$).\\
Given the experimental error bars, it is difficult to clearly delineate which of Ciaravella \textit{et al.}'s interacting model or GT's model fits best the measured data. Actually, to first order, both models fit equally well the experiments, and constitute, to our knowledge, the first direct experimental validation of both models. This suggests in particular, that if one needs to measure the spatial distribution of pressure $p(r)$, GT's model is a very good approximation. Second, it indicates that short range local elastic interactions effects cannot easily be caught when analyzing the radial pressure distribution, or that these effects are of second order.\\
\begin{figure}
	\includegraphics[width=\columnwidth]{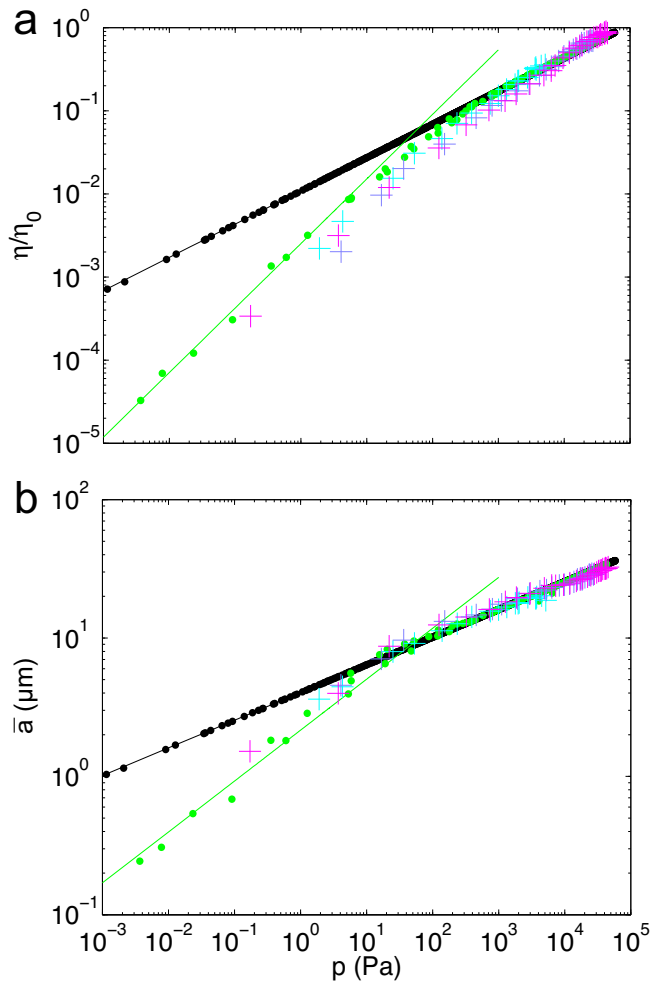}
	\caption{(Color online) (a) Microcontacts density $\eta$, normalized by the mean number of micro-asperities per unit area $\eta_0$, versus local pressure $p$ for the SA sample with $\phi = 0.4$. (b) Mean microcontacts area $\bar a$ versus local pressure $p$ for the same sample. On both graphs, black disks are the results of GT's model predictions, the green disks are predictions of Ciaravella \textit{et al.}'s model with $\alpha_{ij} \neq 0$ and crosses correspond to the experimental data at three different loads $P=0.02, 0.2, 0.5$~N. Thick black lines are power law fits of GT's model predicted data, while green solid lines are power law fits of Ciaravella \textit{et al.}'s model predicted data for $p<p^*$, with $p^* \approx 50$~Pa.}
	\label{fig:naCiaGT} 
\end{figure}
\indent The fact that $p(r)$ distributions are very similar for both models motivates a closer examination of the distributions of quantities from which $p(r)$ derives. For that purpose, the pressure dependence of surface density $\eta$ and mean radius $\overline{a}$ of microcontacts was considered (where $\eta$ is defined as the number of microcontacts per unit area). In Fig~\ref{fig:naCiaGT}, theoretical (as calculated from Ciavarella's model with $\alpha_{ij} \neq 0$) and experimental values of $\eta$ and $\overline{a}$ are reported in a log-log plot as function of the contact pressure $p$. Two different domains are clearly evidenced. When the pressure is greater than a critical value $p^*$, which is here of the order of 50 Pa, $\eta$ and $\overline{a}$ exhibit with $p$ a power law behavior whose exponents are found to be equal to $0.4$ and $0.2$, respectively, from the simulated data. As detailed in Appendix B, these exponents are identical to that predicted by the GW model for nominally flat surfaces in 
the case of a uniform distribution of asperities heights ($\eta \propto p^{2/5}$ and $\overline{a} \propto p^{1/5}$). This means that as long as $p>p^*$, the pressure dependence of $\eta$ and $\overline{a}$ is insensitive to both the effects of the elastic coupling between micro-asperities contacts and to the curvature of the nominal surfaces. Below the critical pressure $p^*$, a power law dependence of $\eta$ and $\overline{a}$ is still observed but with exponents, respectively  $0.78 \pm 0.11$ and $0.37 \pm 0.02$, which depart from the GW predictions (Fig.~\ref{fig:naCiaGT}). We do not yet have a definite explanation for these deviations which are systematically observed, irrespective of the number of surface realizations (up to 8000) considered. They could tentatively be attributed to some short range effects of the pair correlation function associated with asperity distribution. However, the important point is that $p^*$ always corresponds to very low contact pressures. From an extended set of numerical 
simulations where parameters such as asperities density, radius of curvature and height distribution were varied by at least one order of magnitude, $p^*$ was systematically found to be in the range $10^1-10^3$Pa. For the considered contact conditions, such a pressure range corresponds to a very narrow domain at the tail of the pressure distribution whose physical relevance is questionable. In other words, both the simulations and the experimental data indicate that the GW theory is able to describe accurately the microcontacts distribution over most of the investigated pressure range without a need to incorporate the effects of short range elastic interactions in the rough contact description.
\section*{Frictional properties}
\begin{figure}
	\includegraphics[width=\columnwidth]{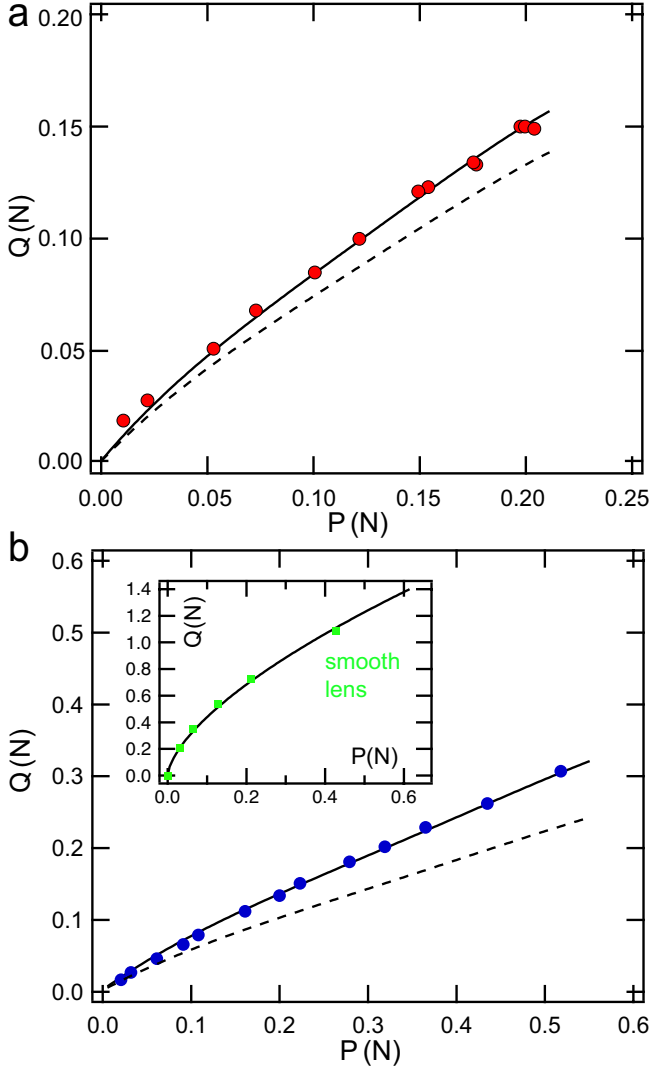}
	\caption{$Q$ versus $P$ in steady sliding ($v=0.5~$mm\,s$^{-1}$) for contacts between a smooth PDMS substrate and RA$^-$ (a) and RA$^+$ (b) lenses. On both graphs, dashed lines are the theoretical $Q$ given by eqn~(\ref{eq:ft}), taking for $A$ its measured values and for $\tau_0=0.34$~MPa the average shear stress obtained with the smooth lens. Solid lines are fits of the experimental data with eqn~(\ref{eq:ft}), yielding $\tau_0=$0.40~MPa for RA$^-$ and 0.49~MPa for RA$^+$. Inset: $Q$ versus $P$ for the smooth lens, in steady sliding. The solid line is a fit of the data using eqn~(\ref{eq:ft}), taking for $A$ its measured value in steady sliding.}
	\label{fig:friction} 
\end{figure}
We now turn onto the frictional behavior of RA lenses against a smooth PDMS slab. As mentioned above, RA asperities are very smooth which allows us to consider the associated micro-asperities contacts as single-asperity contacts. RA surfaces thus provide systems with a single roughness scale as opposed to SA surfaces which present an additional microscopic roughness. In what follows, we address from preliminary results the issue of the contribution of individual micro-asperities contact to the macroscopic friction force. For $P$ within [0.01--0.6]~N and driving velocities $v$ up to 5~mm\,s$^{-1}$, both RA$^+$ and RA$^-$ lenses systematically exhibited smooth steady state friction with no evidence of contact instabilities such as stick-slip, nor strong changes in their frictional behavior. Thus, only results obtained at the intermediate velocity of $v=0.5$~mm\,s$^{-1}$ are reported here. Figure \ref{fig:friction} shows the resulting lateral force $Q$ versus normal force $P$ curves for both RA$^{-}$ (Fig.~\ref{fig:friction}a) and RA$^{+}$ (Fig.~\ref{fig:friction}b) samples, as well as for a reference glass lens with the same radius of curvature and covered with a thin smooth layer of the same sol-gel material used for RA lenses (Fig.~\ref{fig:friction}b, inset). In all cases, $Q$ is found to vary non-linearly with $P$. In the simplest description, the total friction force $Q$ is expected to be the sum of local friction forces $q_i$ acting on all contacting micro-asperities. According to previous studies using glass/PDMS elastomer contacts~\cite{nguyen2011,chateauminois2008}, a constant, pressure independent, shear stress $\tau_0$ can be assumed to  prevail at the intimate contact interface between the asperities and the PDMS elastomer, yielding $q_i=\tau_0 (\pi a_i^2)$. Within this framework, $Q$ should thus write as
\begin{equation}
	Q=\tau_0 A
	\label{eq:ft}
\end{equation}
with $A=\sum_i (\pi a_i^2)$ the real area of contact. In the calculation, we take for $A$ the experimental values measured under normal indentation after verifying from optical contact observations that the microcontacts areas during sliding are not significantly different from that achieved under static loading \footnote[2]{When looking carefully, a slight decrease of individual areas of microcontacts can be seen between the static and sliding regime. This decrease remains however difficult to quantify.}. As a first attempt, the frictional shear stress $\tau_0$ was taken as the experimental value calculated from the ratio of the friction force to the actual contact area measured during steady state friction with the smooth lens. As shown by the dotted lines in Figs.~\ref{fig:friction}a-b, choosing this shear stress value underestimates the experimental data for both small and large size asperities RA samples. Fitting the experimental data with eqn~(\ref{eq:ft}) using a least square method yields however $\tau_0=0.4$ and 0.49~MPa for small and large size asperities respectively. There is thus some evidence of a dependence of the frictional shear stress on the contact length scale, the shear stress at the microcontacts scale being larger  than that at the scale of a millimeter sized contact ($\sim 18\%$ and $\sim 44\%$ increase for RA- and RA+, respectively). Curvatures of the micro-asperity contacts being larger than that of the smooth contact with the glass lens, the increase in $\tau_0$ at small length scales could be attributed to bulk viscoelastic dissipation as a result of the ploughing of the PDMS substrate by the micro-asperities. However, the fact that $Q$ does not vary significantly when the sliding velocity is changed by nearly three orders of magnitude (from 0.01 to 5 mm\,s$^{-1}$) does not support this assumption. This weak contribution of viscoelastic dissipation to friction can be related to the low glass temperature $T_g=-120^{\circ}$C of the PDMS elastomer. Indeed, for the considered micro-asperities size distributions, 
the characteristic strain frequency associated with the microcontacts deformation is $v/a \approx 10$~Hz, \textit{i.e.} well below the glass transition frequency at room temperature (more than $10^8$~Hz). Other effects, arising for example from non linearities in the highly strained microcontacts could be at play, which will be the scope of further investigations. However, these 
experimental results show that frictional stresses measured at macroscopic length scales may not be simply transposed to microscopic multicontact interfaces. 
\section*{Conclusion}
In this work, we have studied both normal contact and friction measurements of model multicontact interfaces formed between smooth surfaces and rough surfaces textured with a statistical distribution of spherical micro-asperities. Two complementary interfacial contacts were studied, namely a rigid sphere covered with rigid asperities against a smooth elastomer, and a smooth rigid sphere against a flat patterned elastomer. In both cases, experimental $A(P)$ relationships were found to be non-linear and well fitted by Ciaravella \textit{et al.}'s model taking into account elastic interaction between asperities. Additional information regarding the nature of the elastic coupling between asperities was provided from the examination of the profiles of contact pressure, contact density and average radius of asperity contacts. While the long range elastic coupling arising from the curved profile of the indenter was found to be an essential ingredient in the description of the rough contacts, both experimental and 
simulation results demonstrate that, for the considered topographies, short range elastic interactions between neighboring asperities does not play any detectable role. As a consequence, the pressure dependence of both the density and the radius of asperity contacts within the macroscopic contact is very accurately described using GW model which neglects asperity interactions. To our best knowledge, these results constitute the first direct experimental validation of GW and GT models. The question arises as to what extent our conclusion regarding the elastic coupling could be extrapolated to more realistic surface roughnesses as theoretical simulations using, for example self affine fractal surfaces, indicate a significant contribution of such effects. From an experimental perspective, this issue could be addressed by considering more sophisticated patterned surfaces with hierarchical distributions of micro-asperities.
\section*{Acknowledgments}
We acknowledge funding from ANR (DYNALO NT09-499845). Many thanks are also due to J.P.~Gong (Hokkaido University, Japan) for her kind support to this study. We are indebted to J.~Teisseire for his support in the fabrication of the sol-gel structures, and to F.~Martin for the SEM images of both RA and SA samples, and thank E.~Barthel for stimulating discussions. V.~Romero is also grateful for a CONICYT financial support from Chile.
\appendix
\section*{Appendix}
\numberwithin{equation}{subsection}
\renewcommand{\thesubsection}{\Alph{subsection}}
\subsection{Gap between surfaces in Hertzian contact}
In a Hertzian sphere--on--flat contact, the vertical displacement $u_z$ of the free surface outside the contact can be expressed as \cite{Johnson1985a}
\begin{equation}
	\begin{split}
		u_z(r)=\frac{4}{3K}\frac{p_0}{2a} \left[ \left(2a^2-r^2 \right)\arcsin(a/r)\right. \\
		\left. +\:ra\left(1-a^2/r^2 \right)^{1/2}\right]\:\:; \: r\geq a
		\label{eq:uz}
	\end{split}
\end{equation}
where $p_0$ is the maximum Hertzian pressure, $a$ is the contact radius and $K$ is the elastic constant defined by $K=4/3 E/(1-\nu^2)$. From the expression of the maximum contact pressure
\begin{equation}
	p_0=\frac{3}{2\pi}\frac{aK}{R_l}
\end{equation}
where $R_l$ is the radius of the spherical indenter, equation~(\ref{eq:uz}) can be rewritten as
\begin{equation}
	\begin{split}
		u_z \left( r \right)=\frac{1}{\pi R_l}\left[ \left(2a^2-r^2 \right)\arcsin(a/r) \right. \\
		\left. +\:ra\left(1-a^2/r^2 \right)^{1/2}\right]\:\:; \: r\geq a
	\end{split}
\end{equation}
The profile of the sphere is given by
\begin{equation}
	s(r)=\frac{1}{2R_l}\left(2a^2-r^2 \right)
\end{equation}
The gap $\left[ u \right] \left( r \right)$ between both surfaces is thus given by
\begin{equation}
	\begin{split}
		\left[ u \right] \left( r \right)=\frac{1}{\pi R_l}\left[ \left(2a^2-r^2 \right)\arcsin(a/r)+ra\left(1-a^2/r^2 \right)^{1/2}\right] \\
		-\frac{1}{2R_l}\left(2a^2-r^2 \right)
	\end{split}
	\label{eq:ur}
\end{equation}
A series expansion of eqn~\ref{eq:ur} at $r=a$ yields
\begin{equation}
	\left[ u \right] \left( r \right) \sim \frac{8}{3}\frac{\sqrt a \sqrt 2}{\pi R_l}\left(r-a \right)^{3/2}+O((r-a)^2)
\end{equation}
For a rough contact, a characteristic length $\Delta$ can be defined as the length over which the above calculated gap between both surfaces is of the order of magnitude of some length characterizing the asperity distribution, like the standard deviation of the height distribution $\sigma$. From the condition $\left[u\right]\left(a+\Delta\right)=\sigma$,
\begin{equation}
	\Delta \simeq \left( \frac{3 \pi}{8 \sqrt 2}\right )^{2/3}\frac{R_l^{2/3} \sigma^{2/3}}{a^{1/3}}
\end{equation}
or
\begin{equation}
	\frac{\Delta}{a} \simeq \left( \frac{3 \pi}{8 \sqrt 2}\right )^{2/3}\frac{R_l^{2/3} \sigma^{2/3}}{a^{4/3}}
\end{equation}
which can also be expressed as a function of the applied normal load $P$
\begin{equation}
	\begin{split}
		\Delta &\simeq \left( \frac{3 \pi}{8 \sqrt 2}\right )^{2/3} R_l^{5/9} \sigma^{2/3} K^{1/9} P^{-1/9}\\
		\frac{\Delta}{a}&\simeq \left( \frac{3 \pi}{8 \sqrt 2}\right )^{2/3} \left( \frac{K^2R_l\sigma^3}{P^2} \right)^{\frac{2}{9}}
	\end{split}
	\label{eq:deltac}
\end{equation}
\subsection{GW's model for a uniform height distribution of spherical asperities}
In this Appendix, we formulate the classical GW's model for the contact between two nominally plane rough surfaces in the case of a uniform height distribution of the spherical asperities. Accordingly, non interacting Hertzian contacts are assumed to occur locally at the scale of the micro-asperities. The surface density of microcontacts is given by
\begin{equation}
	\eta=\int_d^\infty \psi(z)dz
\end{equation}
where $d$ is the separation between the reference planes of the two surfaces and $\psi(z)$ is the expected number of contacts per unit area at a height between $z$ and $z+dz$ above the reference plane. Similarly, the contact pressure $p$ for a given approach $d$ between the surfaces can be defined as
\begin{equation}
	p=\int_d^\infty KR^{1/2}\left(z-d \right )^{3/2}\psi \left(z\right)dz
\end{equation}
where $p$ is defined as the ratio of the applied normal load to the nominal area of contact and $K=4/3 E/(1-\nu^2)$. In the case of a uniform distribution of asperity height with standard deviation $\sigma$, one can write
\begin{equation}
	\int_{-\infty}^\infty \psi(z)dz=k \sigma = \eta_0
\end{equation}
where $k$ is a constant and $\eta_0$ is the surface density of asperities. The surface density of contacts and the contact pressure can then be rewritten as
\begin{equation}
	\eta=\int_0^{\Delta-d} kdx
\end{equation}
\begin{equation}
	p=\int_0^{\Delta-d} KR^{1/2}x^{3/2} dx
\end{equation}
which gives
\begin{equation}
	\eta=k\left( \Delta-d \right)=\frac{\eta_0}{\sigma}\left(  \Delta-d \right)
	\label{eq:eta}
\end{equation}
\begin{equation}
	p=\frac{2}{5} KR^{1/2}\left( \Delta-d \right)^{5/2} \frac{\eta}{\eta_0}
	\label{eq:p1}
\end{equation}
where $\Delta$ is the maximum asperity height above the reference plane. From eqns (\ref{eq:eta}) and (\ref{eq:p1}), the relationship between the surface density of contacts and the contact pressure can be expressed as
\begin{equation}
	\frac{\eta}{\eta_0}=\left ( \frac{5}{2} \right )^{2/5}\left [ \frac{p}{\eta_0KR^{1/2}\sigma^{3/2}} \right ]^{2/5}
	\label{eq:density}
\end{equation}
According to the Hertzian behaviour of micro-asperity contacts, the relationship between the expected mean contact radius $\overline{a}$ and the contact pressure is given by
\begin{equation}
	p=\frac{K}{R}\eta \overline{a}^3
	\label{eq:p2}
\end{equation}
By inserting eqn~(\ref{eq:p2}) in eqn~\ref{eq:density}), the expected mean contact radius may be expressed as
\begin{equation}
	\overline{a}=\left ( \frac{2}{5} \right )^{2/5}\left [ \frac{pR^2\sigma^{2/3}}{K\eta_0} \right ]^{1/5}
	\label{eq:radius}
\end{equation}


	\bibliographystyle{unsrt} 

\end{document}